\documentstyle[12pt]{article}

\def\frac#1#2{{#1\over #2}}

\def\frac#1#2{{#1\over #2}}

\def\ma#1#2#3#4{\left[{}^{#1}_{#3}{}^{#2}_{#4}\right]}

\newcommand{\fAC}{\theta\left[{}_1^1{}_0^0\right]}
\newcommand{\fCA}{\theta\left[{}_0^0{}_1^1\right]}

\newcommand{\fBA}{\theta\left[{}_0^1{}_1^1\right]}

\newcommand{\ffCC}{\theta^2\left[{}_0^0{}_0^0\right]}
\newcommand{\ffCB}{\theta^2\left[{}_0^0{}_0^1\right]}
\newcommand{\ffCO}{\theta^2\left[{}_0^0{}_1^0\right]}
\newcommand{\ffBC}{\theta^2\left[{}_0^1{}_0^0\right]}

\newcommand{\ffAC}{\theta^2\left[{}_1^1{}_0^0\right]}
\newcommand{\ffCA}{\theta^2\left[{}_0^0{}_1^1\right]}
\newcommand{\ffBB}{\theta^2\left[{}_0^1{}_0^1\right]}
\newcommand{\ffBA}{\theta^2\left[{}_0^1{}_1^1\right]}

\newcommand{\ffBO}{\theta^2\left[{}_0^1{}_1^0\right]}

\newcommand{\ffAO}{\theta^2\left[{}_1^1{}_1^0\right]}

\newcommand{\ffOO}{\theta^2\left[{}_1^0{}_1^0\right]}

\newcommand{\kI}{\sqrt{1-\kappa^2}}
\newcommand{\lI}{\sqrt{1-\lambda^2}}
\newcommand{\mI}{\sqrt{1-\mu^2}}

\newcommand{\ml}{\sqrt{\lambda^2-\mu^2}}
\newcommand{\mk}{\sqrt{\kappa^2-\mu^2}}
\newcommand{\lk}{\sqrt{\kappa^2-\lambda^2}}

\begin{document}

\title{Quasiperiodic Solutions of the Fibre Optics Coupled Nonlinear
Schr{\"o}dinger Equations}
\author{P L Christiansen${}^{1}$ , J C Eilbeck${}^{2}$ \\ V Z
Enolskii${}^{2,3}$ , N  A Kostov${}^{4}$ \\
${}^{1}$  Laboratory of Applied Mathematical Physics\\ Danish
Technical University\\  Bld. 303, Lyngby 2800, Denmark \\
${}^{2}$ Department of Mathematics, Heriot-Watt University \\
Riccarton, Edinburgh EH14 4AS, UK \\ ${}^{3}$ Department of Theoretical
Physics\\ Institute of Metal Physics, Vernadsky str. 36\\ Kiev-680,
252142, Ukraine\\
${}^{4}$ Institute of Electronics\\Bulgarian Academy of
Sciences\\ Blvd. Tsarigradsko shosse 72, Sofia 178 4, Bulgaria}
\maketitle

\newpage \begin{center}{\bf Quasiperiodic Solutions of the Fibre
Optics\\ Coupled Nonlinear Schr{\"o}dinger
Equations}\end{center}\vskip 1cm

 \begin{center}P L Christiansen , J C Eilbeck \\
V Z Enolskii, N  A Kostov \end{center} \vskip 2cm

\begin{abstract}
We consider  travelling periodical and quasiperiodical waves in single
mode fibres,  with  weak  birefringence and under the action of
cross-phase  modulation. The problem is reduced to the ``1:2:1"
integrable case  of the two-particle quartic potential. A general
approach for finding elliptic solutions is given.  New solutions which
are associated with  two-gap Treibich-Verdier potentials are found.
General quasiperiodic solutions are given in terms of two dimensional
theta functions with explicit expressions for frequencies in terms of
theta constants. The reduction of  quasiperiodic solutions to elliptic
functions is discussed.
\end{abstract}

\newpage
\section{Introduction}
\setcounter{equation}{0}

We consider the system of two coupled nonlinear Schr\"{o}dinger
equations
\begin{eqnarray}
&&i A_Z + \frac {1} {2}A_{TT}+\sigma
B+(|A|^2+\gamma |B|^2)A =0 ,  \nonumber\\ &&i B_Z + \frac {1} {2}
B_{TT} + \sigma A+(|B|^2 +\gamma |A|^2)B=0.\label{equations}
\end{eqnarray}
These equations are important for a number of physical applications.
For example the dimensionless circularly polarized components $A$, $B$
of the electric field in a single-mode  straight  (no  twists)  fibre
satisfy these equations \cite{bdw87,me87}.  If the functions $A$ and
$B$ do not depend on the variable $t$, these equations are the
generalized discrete self-trapping dimer system \cite{jc93}, which are
integrable in terms of elliptic functions. If $\gamma = 0$ these
equations are the well known DST dimer equations \cite{els85}.  The
equations (\ref{equations}) do  not belong to the class of nonlinear
evolution equations integrable by means of the inverse scattering
method \cite{stl86}.  Nevertheless, they  can have  exact  vector
solitons \cite{cj88,ts88}, bound solitary waves \cite{ch88}  and
periodical solutions  \cite{ft89,kous92}. This interesting phenomenon
is related to the existence of a fourth integral found by Dowling
\cite{do90}.  Due to this property, two-soliton solutions also exist
\cite{ts88}.

In the present paper the general polarization-modulated states --
two-gap solutions -- are expressed in terms of two-dimensional theta
functions. Another  motivation for  this work is the attempt to
classify the exact solutions of the Coupled Nonlinear Schr{\"o}dinger
equation according to the program of classification given in
\cite{mp93}.

Let us introduce the new functions $A=(a+ib) /
\sqrt{2}$, $B=(a-ib) / \sqrt{2}$ and new independent variables
$z=(\gamma + 1 )\,Z / 2$, $t=\sqrt{ \gamma+1 }\,{T}$  \cite{ft89}.
Then the equations (\ref{equations}) can be rewritten as follows
\begin{eqnarray}
i a_z+a_{tt}+\Omega_0 a+p (|a|^2+|b|^2)a+ q(a^2 + b^2)a^{*}&= &0,
\nonumber\\
i b_z+b_{tt}-\Omega_0 b+p (|b|^2 + |a|^2) b+q(a^2+b^2)b^{*}&=&0,
\label{rnse}
\end{eqnarray}
where $\Omega_0 = 2\sigma/3$, $p = \sigma/3$ and $q=1/3$.

We seek solutions of equations (\ref{rnse}) in the  following form
\cite{ft89}
\begin{eqnarray}
a(z,t) = q_1(t) \exp ( i \Omega z ), \quad
b(z,t) = q_2(t) \exp ( i \Omega z ), \label{ansatz}
\end{eqnarray}
where $q_1(t)$ and $q_2(t)$ are real functions, and $\Omega$ is an
arbitrary  real constant.  We obtain the system
\begin{eqnarray}
&&\frac {d^{2}} {dt^{2}} q_1 + ( q_1^2 + q_2^2 ) q_1  =
( \Omega - \Omega_0 ) q_1, \nonumber\\
&&\frac {d^{2}} {dt^{2}} q_2 + ( q_1^2 + q_2^2 ) q_2  =
( \Omega + \Omega_0  ) q_2.\label{121}
\end{eqnarray}

These equations describe the known integrable case ``1:2:1" of the
quartic potential ${\cal U}=Aq_1^4+Bq_1^2q_2^2+Cq_2^4$ \cite{hi87},
which is separable in ellipsoidal coordinates.  The foregoing analysis
can be applied also to the integrable potential ``1:12:16" which is
separable in parabolic coordinates.  The dynamics of the other
separable cases of the quartic potentials ``1:6:1" and ``1:6:8" can be
expressed in terms of two elliptic functions with different moduli
\cite{rrg94}.

The paper is organised as follows. In  Section 2 we describe the
Poisson structure of the integrable system ``1:2:1" using the results
given in \cite{eekl93aa,eekt94}. We derive a Lax representation for the
system, prove its complete integrability and construct the separated
coordinates. In  Section 3 we show that the problem of the description
of elliptic solutions for the system (\ref{121}) is reduced to the
description of the elliptic potentials for the Schr\"odinger equation
and the construction of the corresponding Lam\'e polynomials. The
problem is closely related to the structure of the locus for the
Calogero-Moser system \cite{amm77}. In  Section 4 we analyse the known
elliptic solutions for the system (\ref{121}) and show that they are
included into the approach developed. Moreover we find a new elliptic
solution for the system (\ref{121}) which is associated with the two
gap Treibich-Verdier potential \cite{tv91}. In Section 5 we give an
integration of the system via  theta functions and calculate the
frequencies in terms of theta constants. The question of the periodic
solutions is formulated from a theta functional point of view and
reduced to the description of some modular varieties given by the
vanishing of some two dimensional theta constants. Such a reduction is
demonstrated on the example of the elliptic solution associated with
the Treibich-Verdier potential.

\section{The Lax representation}
\setcounter{equation}{0}

The system (\ref{121}) is a completely integrable Hamiltonian system
with the Hamiltonian
\begin{equation}
H={1\over2}(p_1^2+p_2^2)+{1\over4}(q_1^2+q_2^2)^2-
{1\over2}(z_1q_1^2+z_2q_2^2), \label{h}
\end{equation}
where $(q_i,p_i), \;i=1,2$ are canonical variables with $p_i=\dot
q_i=dq_i/dx, i=1,2$, and  $z_1, z_2$ are arbitrary constants.

The Lax representation for the system (\ref{121}) is known in  terms of
$3\times3$ matrices \cite{zs82}. However, we find it convenient to use
a $2\times2$ Lax representation, $\dot L(z)=[L(z),M(z)]$ which is a
special case of the hierarchy of separable systems discussed in
\cite{eekl93aa,eekt94}.

To do this we fix the following ansatz for the Lax operator.
\begin{eqnarray}
L(z)=\left(\matrix{ V(z)&U(z)\cr
W(z)&-V(z)}\right),\quad M(z)=\left(\matrix{ 0&1\cr
Q(z)&0}\right),\label{Lax}
\end{eqnarray}
where
\begin{eqnarray}
U(z)&=&1 -{1\over2}\sum_{i=1}^2{q_i^2\over z -z_i},\quad
V(z)=-{1\over2}\dot U(z),\label{uz}\\
W(z)&=&{1\over2}\sum_{i=1}^2{p_i^2\over z-z_i}-Q(z), \quad
Q(z)=z+{1\over2}\sum_{i=1}^2 q_i^2.\label{qz}
\end{eqnarray}

The associated algebraic curve ${\rm det}(L(z)-yI)=0$ has genus two,
and is written as
\begin{eqnarray}
 w^2  &=& -4(z-z_1)(z-z_2)(z^3 - z^2 (z_1+z_2)+z(z_{1}z_{2}-H)-F),
\label{curve}\\
&=&-4\prod_{i=1}^5(z-z_i),\nonumber
\end{eqnarray}
where $w=y(z-z_1)(z-z_2)$, $H$ is the
Hamiltonian (\ref{h}), $F$ is the additional integral of motion,
\begin{eqnarray}
F &=& {1\over4}( p_1 q_2 -p_2 q_1 )^2 +{1\over2}(
q_1^{2} + q_2^{2} ) \, \left( z_{1} z_{2} - \frac{z_{2}}{2} q_1^{2}
- \frac{z_{1}}{2 } q_2^{2} \right)\nonumber\\
&&-{1\over2}(z_2p_1^2-z_1p_2^2),\label{f}
\end{eqnarray}
and $z_3, z_4, z_5$ are the roots of the cubic on the rhs.\ of
(\ref{curve}).
Let us define  new coordinates $\mu_1, \mu_2$ as zeros of the
function $U(z)$ in the Lax operator, i.e.
\begin{equation}
q_1^2=-2{(z_1-\mu_1)(z_1-\mu_2)\over z_1-z_2} ,\quad
q_2^2=-2{(z_2-\mu_1)(z_2-\mu_2)\over z_2-z_1}\label{qq}
\end{equation}
On can prove (see e.g. \cite{eekl93aa,eekt94}) that the canonically
conjugated momenta are defined as
\begin{equation}
\pi_i=V(\mu_i)={w_i\over
(\mu_i-z_1)(\mu_i-z_2)};\nonumber
\end{equation}
\begin{equation}
\pi_1={\dot
\mu_1\over2}{\mu_1-\mu_2\over (\mu_1-z_1)(\mu_1-z_2)},\quad \pi_2=
{\dot
\mu_2\over2}{\mu_2-\mu_1\over (\mu_2-z_1)(\mu_2-z_2)} \label{pi}
\end{equation}
and therefore the coordinates $(\mu_i,\pi_i)$ are the
separated coordinates (ellipsoidal coordinates).

It follows from (\ref{qq},\ref{pi}) that the
the dynamics of the system  described in the coordinates
$(\pi_i,\mu_i)$ become  the Jacobi inversion problem associated with
the curve (\ref{curve})
\begin{eqnarray}
\int_{\mu_0}^{\mu_1}{d\mu\over
w(\mu)}+\int_{\mu_0}^{\mu_2}{ d\mu\over
w(\mu)}=a,\nonumber\\ \int_{\mu_0}^{\mu_1}{\mu d\mu\over
w(\mu)}+\int_{\mu_0}^{\mu_2}{\mu d\mu\over
w(\mu)}=2x+b,\label{jip}
\end{eqnarray}
where $a,b$ are constants defined by the initial conditions. If
$z_{1,2}$ are real and $z_1>z_2$ then the solution (\ref{qq}) is real
if $\mu_2\leq z_1\leq\mu_1$ and $\mu_1\leq z_2$ or $z_2\leq \mu_2$.

We remark that the integrable case ``1:12:16" of the quartic potential
arises in this approach by the introduction of another ansatz for the
function $U$ whose zeros define parabolic coordinates
\cite{eekl93aa}.

\section{Periodic solutions associated with the Lam\'e equation}
\setcounter{equation}{0}

In this section we develop a method (see also
\cite{ee94b,ek93,kost89}) which allows us to construct periodic
solutions   of (\ref{121})  in a straightforward way.  The method is
based on  the application  of spectral theory for the Lam\'e equation
with elliptic potentials \cite{amm77,mm75}.
\begin{eqnarray} \frac
{d^{2}} {dx^{2}} \Psi(x,z) - {\cal U}(x) \Psi(x,z) = -z \Psi(x,z).
\label{lame}
\end{eqnarray}
with ${\cal U}(x)$ being an elliptic potential.

Because the algebraic curve (\ref{curve}) associated with the problem
has genus two we shall consider a two-gap elliptic potential for the
equation (\ref{lame}). Such  potentials are known to be of the
form \cite{amm77}
 \begin{equation}
{\cal U}(x)=2\sum_{i=1}^N\wp(x-x_i),\label{elpot}
\end{equation}
where $\wp(x) $ is the Weierstrass elliptic functions \cite{ba55} with
the real period $2\omega=2\omega_1$ and imaginary period
$2\omega'=2\omega_3$. The number $N$ is a positive integer $N>2$ (the
number of ``particles") and the numbers ${\bf x}=(\tilde
\mu_1,\ldots,x_N)$ belongs to the locus ${\cal L}_N$, i.e., the
geometrical position of the points given by the equations
\begin{equation}
{\cal L}_N=\left\{({\bf x});\sum_{i\neq
j}\wp'(x_i-x_j)=0,\; j=1,\ldots N)\right\}.\label{locus}
\end{equation}
Equation
(\ref{lame}) allows the coalescence of three particles $x_i$ and the
potential takes the form \cite{amm77}.
\begin{equation}
{\cal U}(x)=6\sum_{i=1}^n\wp(x-x_i)+2\sum_{j=1}^m\wp(x-x_j),\quad
3n+m=N\label{elpot1}
\end{equation}

The associated algebraic curve of genus two can be described with the
help of the Novikov equation \cite{no74}. For example, let us consider
the two-gap potential for (\ref{elpot1}) normalized by its expansion
near the singular point as
\begin{equation}
{\cal U}(x) = {6\over x^2} + a x^2 + b
x^4 + c x^6 + d x^8 + O(x^{10}).  \label{decomposition}
\end{equation}
Then, following from the Novikov equation \cite{no74}, the algebraic
curve associated with this potential has the form \cite{be89b}
\begin{equation}
w^2=
  z^5 - {{35 a {z^3}}\over 2}  - {63 b {z^2}\over 2}+ \left( {{567
  {a^2}}\over 8} + {{297 c}\over 4} \right) z+{{1377 a b}\over 4} -
  {{1287 d}\over 2}. \label{curvelame}
\end{equation}

Let us consider the {\it trace formulae} \cite{zmnp80} written for the
elliptic potential in the form
\begin{eqnarray}
\mu_1+\mu_2 &=&-\sum_{j=1}^N \wp (x - x_j) +
{1\over 2} \sum_{j=1}^5 z_j,\label{tr1} \cr
\mu_1 \mu_2 &=&3\,\sum_{i<j} \wp(x - x_i) \wp(x -x_j) -
{Ng_2\over8} \nonumber\\
&&+{1\over 2} \sum_{i<j} z_i z_j - {3\over
8}\left(\sum_{j=1}^5 z_j \right)^2 \label{tr2}
\end{eqnarray}

We require that the eigenvalues $z_1$ and $z_2$  be the branching
points of the curve associated with the two-gap elliptic potential.
Using the relation which follows from (\ref{qq})
\begin{equation}
q_1^2+q_2^2=2(\mu_1+\mu_2-z_1-z_2),\label{q+q}
\end{equation}
then from the first  trace formula we can write (\ref{121}) in the
form
\begin{eqnarray}
&&\frac {d^{2}} {dx^{2}} q_1 -{\cal U}(x) q_1 = (
\Omega - \Omega_0 + 2 z_{1} + 2 z_{2} ) q_1 \\ &&\frac {d^{2}} {dx^{2}}
q_2 -{\cal U}(x) q_2 = ( \Omega + \Omega_0 + 2 z_{1} + 2 z_{2} ) q_2,
\end{eqnarray}
where we set without  loss of generality $\sum_{j=1}^5 z_j=0$.

Then if the relations
\begin{equation}
\Omega - \Omega_0 = - 3 z_{1} -
2 z_{2}, \quad \Omega + \Omega_0  =-2z_{1}-3
z_{2}.\label{dispersion}
\end{equation}
hold,  the problem of finding  elliptic solutions for (\ref{121}) is
reduced to the calculation of two Lam\'e polynomials, i.e. the values
of the eigenfunctions of (\ref{Lax}) corresponding to the spectral
parameter fixed at the ends of the gaps
\begin{equation}
q_1=C_1\Psi(x,z_1),\quad
q_2=C_2\Psi(x,z_2),\label{lamepol}
\end{equation}
where $C_i$ are some constants, corresponding to a point ${\bf x}$
on the locus ${\cal L}_N$.

For the elliptic potential without degeneracy (\ref{elpot}) we have
from (\ref{qq},\ref{tr1}) and (\ref{q+q}) the general formula
for elliptic solutions of  equations (\ref{121}).
\begin{eqnarray}
q_1^2&=&{1\over z_2-z_1}\left(2z_1^2+2z_1\sum_{i=1}^N\wp(x-x_i)
\right.\nonumber\\
&&+\left.6\sum_{1\leq i< j\leq N}\wp(x-x_i)\wp(x-x_j)-{Ng_2\over4}+
\sum_{1\leq i< j\leq 5}z_iz_j\right),\label{q1}\\
q_2^2&=&{1\over z_1-z_2}\left(2z_2^2+2z_2\sum_{i=1}^N\wp(x-x_i)
\right.\nonumber\\
&&+\left.6\sum_{1\leq i< j\leq N}\wp(x-x_i)\wp(x-x_j)-
{Ng_2\over4}+\sum_{1\leq i< j\leq 5}z_iz_j\right).\label{q2}
\end{eqnarray}

A problem with the application of this formulae is that we have to
find  points ${\bf x}$ on the locus such that the functions
(\ref{q1},\ref{q2}) are real and finite. Some special cases are known,
for example the analytical description of the locus (\ref{locus}) for
two-gap elliptic potentials for which $x_i$ are half-periods,
$x_i=\omega_i$ (Treibich-Verdier potentials, \cite{tv91}).  Details are
given in \cite{ee94c}.  The curves and Lam\'e polynomial associated
with these potentials are given in  Table \ref{T1}.

\begin{table}[htpb] \caption{The algebraic curves and Lam\'e
polynomials}
\label{T1}\vskip 3mm
\begin{tabular}{||l|l||}\hline
${\cal U}_N$&The curves and
Lame polynomials $\Lambda(x)$ \\ \hline
${\cal U}_3$
&$\rule[-3mm]{0mm}{8mm}
{\cal U}(x)=6\wp(x)$\\
{}&$w^2=-(z^2-3g_2)\prod_{i=1}^3(z-3e_i)
$ \\ {}&$ \Lambda_{ij}=\sqrt{(\wp(x)-e_i)(\wp(x)-e_j)}, \quad (z=3e_k),
\quad i\neq j\neq k=1,2,3
$\\{}&$\rule[-3mm]{0mm}{8mm}\Lambda_{\pm}=\wp(x)\pm {1\over
2}\sqrt{g_2\over3},(z=\pm\sqrt{3g_2})$\\ \cline{2-2} \hline
${\cal U}_4$
&$\rule[-3mm]{0mm}{8mm}
{\cal U}(x)=6\wp(x)+2\wp(x+\omega_i)$\\
{}&$w^2=-(z+6e_i)\prod_{k=1}^4(z-z_k(i)),\quad i=1,2,3$\\{}&
 $z_{1,2}(i)=e_j+2e_i\pm2\sqrt{(e_i-e_j)(2e_j+7e_i)}$\\
{}&$z_{3,4}(i)=e_k+2e_i\pm2\sqrt{(e_i-e_k)(2e_k+7e_i)}$ \\
{}&$\Lambda_{ik}=\sqrt{(\wp(x)-e_i)(\wp(x)-e_j)}+{1\over3}[(e_i-e_k)
\pm\sqrt{(e_i-e_k)(7e_i+2e_k)}]\sqrt{\wp(x)-e_j\over \wp(x)-e_i}$\\
{}&\quad $(z=e_k+2e_i\pm2\sqrt{(e_i-e_k)(2e_k+7e_i)})$ \\
{}&$\rule[-3mm]{0mm}{8mm}\Lambda_{0}=\wp(x)-e_i\;(z=-6e_i)$\\ \cline{2-2}
\hline
${\cal U}_5$ &$\rule[-3mm]{0mm}{8mm}
{\cal U}(x)=6\wp(x)+2\wp(x+\omega_i)+2\wp(x+\omega_j)$\\
{}&$w^2=\prod_{i=1}^{i=5}(z-z_i(k)), \quad
j=1,2,3$\\ {}&$z_4(k)=6e_k-3e_i,\;z_5(k)=6e_i-3e_k$ \\
{}&$\prod_{i=1}^3(z-z_i(k))=z^3-3z^2e_k+(51e_k^2-20g_2)
z-369e_k^3+132e_kg_2$\\
&$\Lambda_i=\sqrt{(\wp(x)-e_i)(\wp(x)-e_k)}+(e_i-e_j)
{\sqrt{\wp(x)-e_k\over\wp(x)-e_i}},\quad(z=6e_i-3e_j)$\\
&$\Lambda_j=\sqrt{(\wp(x)-e_j)(\wp(x)-e_k)}+
(e_j-e_i){\sqrt{\wp(x)-e_k\over\wp(x)-e_j}},\quad (z=6e_j-3e_i)$\\
{}&$\Lambda_n=\sqrt{(\wp(x)-e_i)(\wp(x)-e_j)}+ \tilde a_{ij}
{\sqrt{\wp(x)-e_i\over\wp(x)-e_j}}+\tilde a_{ji}
{\sqrt{\wp(x)-e_j\over\wp(x)-e_i}}, $\\
{}&$\rule[-3mm]{0mm}{8mm}\quad (z=z_n), i\neq j\neq k,\;n =1,2,3,
\tilde a_{ij}
={15e_i^2+27e_j^2-6e_ie_j-z_n^2+2z_n(e_j-e_i)\over
24(e_j-e_i)}$\\\hline \end{tabular}\end{table}

\section{ Elliptic solitary waves}
\setcounter{equation}{0}

We shall show in this Section that the known periodic solutions of
 (\ref{121}) are associated with one and two-gap elliptic
potentials of the Schr\"odinger  equation.

\subsection{One-gap potentials}
The equations (\ref{121}) have the
following solutions \cite{ch88}:
\begin{eqnarray}
q_1 =C_1 {\rm
sn}(\alpha x, k), \quad q_2 = C_2 {\rm cn}(\alpha x, k),
\label{jacfun}
\end{eqnarray}
where the amplitudes $C_1$ ,$C_2$, the modulus $k$ of the elliptic
functions, and the temporal pulsewidth $1/\alpha$ of the waves are
defined in terms of the
physical parameters $\Omega$ and $\Omega_0$ as
\begin{eqnarray}
\alpha^2 k^2 = 2
\Omega_0, \, C^{2}_{1} = \Omega - 3\Omega_0 + \alpha^2  , \,
C^{2}_{2} = \Omega + \Omega_0 + \alpha^2. \nonumber
\end{eqnarray}
The parameters can be expressed in terms of the Weierstrass parameters
$e_3\leq e_2\leq e_1$ as follows, $\Omega=-(e_2+e_3)/2,\;
\Omega_0=(e_2-e_3)/2,\; \alpha =\sqrt{e_1-e_3},\;
C_1=\sqrt{e_1-2e_2},\;
C_2=\sqrt{e_1-2e_3}$

The solutions  (\ref{jacfun}) are  associated  with  the eigenvalues
$z_1 = e_2$ and $z_2 = e_3$  of the one-gap Lam\'e potential. In this
case $H={7\over4}(e_2+e_3)^2+e_2e_2$, $F={1\over2}(e_2+e_3)^2$ and the
curve (\ref{curve}) reduces (after a transformation of the spectral
parameter $z\rightarrow -z+e_2+e_3$) to the product of the Weierstrass
cubic and a perfect square
\begin{equation}
w^2=4(z-e_1)(z-e_2)(z-e_3)\left(z-{3\over2}e_1\right)^2.
\label{Wei}
\end{equation}
It is easy to see that one of the variables
$\mu_i$ is pinched at the point $3e_1/2$ and the Jacobi inversion
problem written for the curve (\ref{Wei}) becomes the inversion of
the elliptic integral for the remaining variable $\mu$.

\subsection{Periodic solutions associated with the two-gap Lam\'e
potential}

It was shown  in \cite{ft89} by direct substitution that the equations
(\ref{121}) have the following periodical solutions:
\begin{eqnarray}
q_1 = C {\rm dn}(\alpha x, k) {\rm sn}(\alpha x, k), \quad
q_2 = C {\rm dn}(\alpha x ,k ) {\rm cn}(\alpha x, k),\label{ft}
\end{eqnarray}
where ${\rm sn}$, ${\rm cn}$, ${\rm dn}$  are  the  standard  Jacobian
elliptic  functions \cite{ba55}, $k$ is the modulus of the elliptic
functions $0 < k < 1$,  and the characteristic  parameters of the wave:
amplitude $C$,   temporal pulsewidth $1/\alpha$ and $k$ are related to
the physical parameters $\Omega$ and, $\Omega_0$ through the following
dispersion relations
\begin{eqnarray}
        C^2 = \frac{6 \Omega} {5} + 2 \Omega_0 , \quad
        k^2 = \frac{10 \Omega_0} {3 \Omega + 5 \Omega_0 } , \quad
      \alpha^2 = \frac {3 \Omega+5 \Omega_0} {15} . \label{dr1}
\end{eqnarray}
Another elliptic solution was found in \cite{kous92}.
\begin{eqnarray}
q_1 = C_1 \alpha k{\rm cn}( \alpha x, k ){\rm sn}( \alpha x, k),
q_2 = C_2{\rm sn}^2 (\alpha x, k) + C_{3},\label{kous}
\end{eqnarray}
where $C$, $C_1$, $\alpha$  and $k$ are expressed through physical
parameters $\Omega$ and $\Omega_0$ by the following relations
\begin{eqnarray}
C_1^{2}&=&\frac {6c}{5},\quad
C_3=-\left(1+\frac{c}{5}\right)\sqrt{\Omega_0},\quad
C_2^{2}=\frac{12\Omega}{5},\nonumber\\
\alpha^{2}&=&\frac {\Omega_0}{2}+\frac {c\Omega_0}{15}-
\frac {\Omega}{10},\quad k^{2}=-\frac{20c}{(c-15)(c+5)},\label{dr2}
\end{eqnarray}
where $c^{2}= 15\Omega/\Omega_0$.

The solutions (\ref{ft},\ref{kous}) can be obtained from the general
periodic solution (\ref{q1},\ref{q2}).  The solutions (\ref{ft}) have
the following spectral interpretation. They are the Lam\'e polynomials
associated with the eigenvalues $3 e_2$, $3 e_3$ (see  Table \ref{T1}),
\begin{eqnarray} q_1^2&=&-{6\over
e_1-e_3}(\wp(x+\omega_3)-e_1)(\wp(x+\omega_3)-e_3)\nonumber\\
q_2^2&=&{6\over
e_1-e_3}(\wp(x+\omega_3)-e_1)(\wp(x+\omega_3)-e_2).\nonumber
\end{eqnarray}
One can show that the relations (\ref{dispersion})
become  (\ref{dr1}).  Analogously one can prove that the normalised
Lam\'e polynomials associated with the eigenvalues  $3 e_2$, $- \sqrt
{3 g_2}$ gives the solution (\ref{kous}).  We prolong this analogue by
constructing some new periodic
solutions corresponding to the Treibich-Verdier potentials given in
Table \ref{T1}.

\subsection{Periodic solutions associated with the two-gap
     Treibich-Verdier potentials}
Below we construct the two periodic solutions associated with the
Treibich-Verdier potential ${\cal U}_3$ in the same way as the
solutions (\ref{ft}) and (\ref{kous}) are connected with the two-gap
Lam\'e potential ${\cal U}_4$. Let us consider the potential
\begin{equation}
{\cal U}_4(x)=6\wp(x+\omega_3)+2{(e_1-e_2)(e_1-e_3)\over
\wp(x+\omega_3)-e_1}\label{tv4}
\end{equation}
and construct the
solution in terms of Lam\'e polynomials associated with the eigenvalues
$z_1,z_2$,  $z_1>z_2$
\begin{eqnarray}
z_1&=&e_2+2e_1+2\sqrt{(e_1-e_2)(7e_1+2e_2)},\nonumber\\
z_2&=&e_3+2e_1+2\sqrt{(e_1-e_3)(7e_1+2e_3)}.\label{zz}
\end{eqnarray}
The finite and real solutions $q_1,q_2$ have the form
\begin{eqnarray}
q_1&=&i  C\sqrt{\wp(x+\omega_3)-e_3}
\left(\sqrt{(\wp(x+\omega_3)-e_1}\right.\nonumber\\&&\hskip 1cm \left.
+{e_1-e_2+\sqrt{(e_1-e_2)(7e_1+2e_2)}\over
3\sqrt{(\wp(x+\omega_3)-e_1}}\right),\label{ee1}\\
q_2&=&C\sqrt{\wp(x+\omega_3)-e_2}
\left(\sqrt{(\wp(x+\omega_3)-e_1}\right.\nonumber\\&&\hskip 1cm \left.
+{e_1-e_3+\sqrt{(e_1-e_3)(7e_1+2e_3)}\over
3\sqrt{(\wp(x+\omega_3)-e_1}}\right).\label{ee2}
\end{eqnarray}
Here $\omega_3$ is the pure imaginary half period, $0\leq t\leq
2\omega$ with $2\omega$ being the real period, and
\begin{equation}
C^2={18\over z_1-z_2}>0.\label{ampl}
\end{equation}
The parameters $\Omega$ and $\Omega_0$ are linked with the Weierstrass
parameters $e_i$ through
\begin{equation}
2\Omega=-5(z_1+z_2),\quad 2\Omega_0=z_1-z_2\label{drtv4}
\end{equation}
with $z_1,z_2$ given by (\ref{zz}). By eliminating $e_i$ from these
formula and from the formula for the amplitude (\ref{ampl}) we arrive
at the dispersion relations,
\begin{eqnarray}
C^2\Omega_0&=&9,\quad 0\leq k^2\leq 1,\nonumber\\ {\Omega\over
5\Omega_0}&=&{k^2-2-2\sqrt{1-k^2}\sqrt{4-k^2} -2\sqrt{4-3k^2}\over
k^2+2\sqrt{1-k^2}\sqrt{4-k^2}-2\sqrt{4-3k^2}},\quad \Omega_0\leq
{\Omega\over 15}.\label{drtv4b}
\end{eqnarray}

Analogously we can find the elliptic solution associated with the
eigenvalues
\begin{eqnarray}
z_1&=&e_2+2e_1+2\sqrt{(e_1-e_2)(7e_1+2e_2)},\quad
z_2=-6e_1,\label{zz2}\end{eqnarray}
We have \begin{eqnarray}q_1&=&C(\wp(x+\omega_3)-e_1),\label{ee3}\\
q_2&=&i C\sqrt{\wp(x+\omega_3)-e_3}
\left(\sqrt{(\wp(x+\omega_3)-e_1}\right.\nonumber\\&&\hskip 1cm \left.
+{e_1-e_2+\sqrt{(e_1-e_2)(7e_1+2e_2)}\over
3\sqrt{(\wp(x+\omega_3)-e_1}}\right),\label{ee4}
\end{eqnarray}
where $C$
is given by (\ref{ampl}) but with $z_1,z_2$ given by (\ref{zz2}). The
corresponding dispersion relation has the form
\begin{eqnarray}
C^2\Omega_0&=&9,\quad {7\over8}\leq k^2\leq 1,\nonumber\\ {\Omega\over
5\Omega_0}&=&{3-2k^2-2\sqrt{1-k^2}\sqrt{4-k^2}\over
5-2k^2+2\sqrt{1-k^2}\sqrt{4-k^2}},\quad 0\leq
{\Omega\over 5\Omega_0}\leq {1\over 3}.  \label{drtv4a}
\end{eqnarray}

The Treibich-Verdier potential ${\cal U}_5$ also yields elliptic
solutions, but the solutions corresponding to the potential ${\cal
U}_5$ given in the Table blow up. From general considerations we
conjecture that there exist non blow-up real solutions associated with
the isospectral deformation of this potential.

The solutions derived  represent stationary periodical waves. However,
taking into account the invariance of  the equations (\ref{equations})
under  a  Galilean transformation  \cite{ft89,do90}, they also
represent travelling periodical waves.

\section{Exact quasi-periodic solutions}
\setcounter{equation}{0}

In this section we give the theta functional expressions for the
trajectories of the  system under consideration using the Rosenhain
ultraelliptic functions \cite{ro51} (see also \cite{mu83}),
i.e.\ Abelian functions associated with an algebraic curve of genus
two. We also show how to reduce ultraelliptic solutions to the elliptic
solutions discussed before.

\subsection{Theta functional integration}

Let  $(z_{\alpha},z_{\beta})\in (z_1,\ldots,z_5)$ be two arbitrary
branching points of the curve (\ref{curve}). After the transformation
$(w,z)\rightarrow(\zeta,\xi)$,
\begin{eqnarray}
w&=&2i(z_{\beta}-z_{\alpha})\sqrt{(z_i-z_{\alpha})
(z_j-z_{\alpha})(z_k-z_{\alpha})}\zeta,\nonumber\\
z&=&(z_{\beta}-z_{\alpha})\xi+z_{\alpha}\label{tr}
\end{eqnarray}
the curve becomes \cite{ro51}
\begin{equation}
\zeta^2=\xi(1-\xi)(1-\kappa^2\xi)(1-\lambda^2\xi)
(1-\mu^2\xi)\label{rcurve}
\end{equation}
with
\begin{equation}
\kappa^2={z_{\beta}-z_{\alpha}\over z_i-z_{\alpha}},\,
\lambda^2={z_{\beta}-z_{\alpha}\over z_j-z_{\alpha}},\,\mu^2=
{z_{\beta}-z_{\alpha}\over z_k-z_{\alpha}}.
\end{equation}

Let us fix on (\ref{curve}) the homology basis, $({\bf
A;B})=(A_1,A_2;B_1,B_2)$ and the conjugated basis of differentials of
the first kind
\[{\bf v}=(v_1,v_2), \quad v_1={c_{11}\xi+c_{12}\over \zeta}d\xi\quad
v_2={c_{21}\xi+c_{22}\over \zeta}d\xi
\]
normalised as
\begin{equation}
\left(\oint_{A_1}{\bf v},\oint_{A_2}{\bf
v};\oint_{B_1}{\bf v},\oint_{B_2}{\bf v}\right)=({\bf 1}_2;\tau),
\end{equation}
where ${\bf 1}_2$ is the $2\times 2$ unit matrix and the period matrix
$\tau$ belongs to the Siegel upper half space ${\cal S}_2$ of degree
$2$. The Jacobi inversion problem can be rewritten as
\begin{eqnarray}
&&\int_{\mu_0}^{\mu_1}v_1+\int_{\mu_0}^{\mu_2}v_1=
2c_{11}\sqrt {z_{\beta}-z_{\alpha}\over \kappa\lambda\mu}x=u_1x+u_{10},\\
&&\int_{\mu_0}^{\mu_1}v_2+\int_{\mu_0}^{\mu_2}v_2=
2c_{21}\sqrt{z_{\beta}-z_{\alpha}\over \kappa\lambda\mu}x=u_2x+u_{20}
\end{eqnarray}
The solution of the problem is expessed in terms of theta functions
with characteristics $[\varepsilon]$ defined on $C^2\times{\cal S}_2$
\begin{equation}
\theta\left[{}^{\varepsilon'}_{\varepsilon''}\right] ({\bf w}|\tau)=
\sum_{n\in
{\bf Z}^2} \exp\left\{\pi i \left<(n+{\varepsilon'\over 2})\tau,
(n+{\varepsilon'\over 2})\right>
+2\pi i\left<n+{\varepsilon'\over 2},{\bf w}+{\varepsilon''\over
2}\right>\right\}\label{theta},
\end{equation}
where $\left<,\right>$ is a Euclidean scalar product and ${\rm Im}\;
\tau $
is positive definite.  For integer characteristics we have
\begin{eqnarray}
\theta\left[{}^{\varepsilon^{\prime}}_{\varepsilon^{\prime\prime}}
\right]
({\bf w}|\tau)&=& {\rm exp}\,\pi\,i\,\left[
{1\over4}\langle\varepsilon^{\prime},\tau\varepsilon^{\prime}\rangle
+\langle \varepsilon^{\prime}{\bf w}\rangle+
{1\over2}\langle\varepsilon^{\prime},\varepsilon^{\prime\prime} \rangle
\right]\nonumber\\&\times&\theta\left({\bf
w}+I{\varepsilon^{\prime\prime}\over2}+ \tau
{\varepsilon^{\prime}\over2}|\tau\right),\label{shift} \end{eqnarray}
where $\theta(z|\tau)=\theta\left[{}_0^0\right](z|\tau)$.
The characteristic $[\varepsilon]$ is called  {\it even}
if the corresponding theta function is an even function
($\langle\varepsilon,\varepsilon'\rangle=0\,({\rm mod}2)$) and
{\it odd} if the corresponding theta function is an odd function
($\langle\varepsilon,\varepsilon'\rangle=1\,({\rm mod}2)$).

The values of theta functions and their derivatives at the  point
$z=0$ are called theta constants which we denote by
$\theta[\varepsilon](0;\tau)=\theta[\varepsilon]$, if the
characteristic $[\varepsilon]$ is even and
$\partial\theta[\varepsilon](z;\tau)/\partial
z_i\big|_{z=0}=\theta_i[\varepsilon],i=1,2$ if the chacteristic
$[\varepsilon]$ is odd.

The function (\ref{theta}) satisfies two sets of
functional equations \cite{mu83}, the {\it transformational property}
\begin{eqnarray}
&&\theta[\varepsilon]({\bf w}+{\bf n}''+{\bf
n}'{\tau}|{\tau}) \nonumber\\&&= \exp \pi i\Bigl[ -\langle {\bf
n}'{\tau},{\bf n}'\rangle - 2\langle {\bf n}'' , {\bf w}\rangle
  +\langle \varepsilon',{\bf n}'\rangle - \langle\varepsilon'',{\bf n}'
\rangle\Bigr]\,\theta[\varepsilon]({\bf w}|{\tau}) \label{trans}
\end{eqnarray}
where ${\bf n', n''} \in {\bf Z}^g$ and the {\it
modular property}, which describes the transformation of the theta
function under the action of the group $Sp_{4}({\bf Z})$.

The branching points are expressed in terms of theta constants as
\begin{eqnarray}
\kappa^2={\ffBB\ffBC\over\ffCB\ffCC},\quad\lambda^2=
{\ffBO\ffBB\over\ffCO\ffCB},\quad \mu^2={\ffBO\ffBC\over\ffCO\ffCC}.
\nonumber
\end{eqnarray}

The Jacobi inversion problem has the solution
\begin{eqnarray}
&&{\ffAO({\bf u} (x-x_0)+{\bf u}_0|\tau)\over\ffOO({\bf u}
(x-x_0)+{\bf u}_0|\tau)}=-\kappa\lambda\mu \; \tilde \mu_1\tilde
\mu_2,\label{r1}\\
&&{\ffBO({\bf u} (x-x_0)+{\bf u}_0|\tau)\over\ffOO({\bf u} (x-x_0)+{\bf
u}_0|\tau)}=-{\kappa\lambda\mu\;(1-\tilde \mu_1)(1-\tilde
\mu_2), \label{r2}\over \kI\lI\mI}\\
&&{\ffCA({\bf u} (x-x_0)+{\bf u}_0|\tau)\over\ffOO({\bf u} (x-x_0)+
{\bf u}_0|\tau)}={\lambda\mu
\;(1-\kappa^2\tilde \mu_1)(1-\kappa^2\tilde \mu_2)\over \kI\lk\mk},
\label{r3}\\
&&{\ffCB({\bf u} (x-x_0)+{\bf u}_0|\tau)\over\ffOO({\bf u} (x-x_0)+
{\bf u}_0|\tau)}={\kappa\mu
\;(1-\lambda^2\tilde \mu_1)(1-\lambda^2\tilde \mu_2)\over
\lI\ml\lk},\label{r4}\\
&&{\ffCC({\bf u} (x-x_0)+{\bf u}_0|\tau)\over\ffOO({\bf u} (x-x_0)+
{\bf u}_0|\tau)}={\kappa\lambda
\;(1-\mu^2 \tilde \mu_1)(1-\mu^2\tilde \mu_2)\over
\mI\mk\ml}.\label{r5}\end{eqnarray}

The solutions of  equations (\ref{121}) are expressed in terms of the
formulae (\ref{r1}-\ref{r5}). To do this we have to put $z_1,z_2$ into
correspondence with two branching points of the curve (\ref{rcurve}).
In particular, let us choose $z_{1,2}$ in such a way that their images
$\tilde z_{1,2}$ under the transformation (\ref{tr}) becomes  the
branching points $0$ and $\lambda^2$ and the images $\tilde \mu_{1,2}$
of $\mu_{1,2}$ move inside the gaps $\tilde z_2=0\leq \tilde\mu_2\leq
\mu^2$, $\tilde z_1=\lambda^2\leq \tilde\mu_1\leq 1$. Let us also fix
the vector ${\bf u}_0$ as a half period ${\bf
u}_0=(1/2)(\tau_{11}+\tau_{12},\tau_{12}+\tau_{22}+1)$, which shifts
the characteristics at $\left[{}_1^1{}_0^1\right]$.  Then we  find the
following quasiperiodic solutions
\begin{eqnarray}
&&
q_1^2
=-2{(z_{\beta}-z_{\alpha})^2\lI\ml\lk\over(z_1-z_2)\kappa\mu}
\nonumber\\
&&\quad \times {\ffAC({\bf
u}(x-x_0)+{\bf u}_0|\tau)\over\ffBA({\bf u} (x-x_0)+
{\bf u}_0|\tau)}\label{qq1}\\
&&q_2^2
=-2{(z_{\beta}-z_{\alpha})^2\over(z_1-z_2)\kappa\lambda\mu}
{\ffCA({\bf u} (x-x_0)+{\bf
u}_0|\tau)\over\ffBA({\bf u} (x-x_0)+{\bf u}_0|\tau)},\label{qq2}
\end{eqnarray}
where the frequences $u_{1,2}$ are in general noncommensurable,
\begin{eqnarray}
u_{1} = \frac{ \sqrt{z_{\beta}-z_{\alpha} }  \theta_{2}
\left[ {}^1_0 {}^1_1  \right]}
            { \pi^{2} \theta\left[ {}^1_0 {}^1_0  \right]
                   \theta\left[ {}^1_0 {}^0_0  \right]
                   \theta\left[ {}^1_0 {}^0_1  \right]},\quad
u_{2} = -\frac{ \sqrt{z_{\beta}-z_{\alpha} }  \theta_{1}
\left[ {}^1_0 {}^1_1  \right]}
            { \pi^{2} \theta\left[ {}^1_0 {}^1_0  \right]
                   \theta\left[ {}^1_0 {}^0_0  \right]
                   \theta\left[ {}^1_0 {}^0_1  \right]},\label{wind}
\end{eqnarray}

We emphasise that the solution given is parametrised only by the the
$\tau$-matrix and expressed in terms of theta functions and theta
constants which are rapidly convergent and therefore this expression is
convenient for numerical calculations.

\subsection{Reduction to elliptic functions}
The given solutions are quasiperiodical functions of time. They are
reduced to elliptic functions under some restrictions on the parameters
of the system which can be formulated in terms of conditions on the
matrix $\tau$.

More precisely, it follows from the explicit formulae
(\ref{qq1},\ref{qq2}) for $q_1,q_2$ that the following two conditions
are sufficient to require the solution be an elliptic function of $x$.

\begin{enumerate}
\item The matrix $\tau$ is taken in the form
\begin{equation}
\tau=\left(\begin{array}{cc}\tau_{11}&{1\over N}\\{1\over
N}&\tau_{22}\end{array}\right),\quad N\in \bf N\label{weierstrass}
\end{equation}

\item $u_1$ or $u_2=0$.
\end{enumerate}

It follows immediately  from the transformational properties of theta
functions that under these conditions the solution is an elliptic
function of the $N$-th order. The condition 2 is reduced with the help
of the formulae (\ref{wind}) to the condition of the vanishing of some
theta constants. It is remarkable that the conditions 1, 2 are also
 sufficient conditions, as was proved in \cite{ee94c} within the
Weierstrass reduction theory of theta functions to lower genera
\cite{bbeim94,kw15}. If only the condition 1 is satisfied then the
solution  (\ref{qq1},\ref{qq2}) is quasiperiodic and expressed in term
of two elliptic functions with the Jacobian moduli $N\tau_{11}$ and
$N\tau_{22}$.

To demonstrate how this approach works we derive the elliptic solution
(\ref{ee1},\ref{ee2}) from the theta functional approach. To this
end we fix the period matrix in the form (\ref{weierstrass}) and
put $N=4$. We remark that this case of the reduction of theta functions
was studied by Bolza \cite{bo87} (see also \cite{bbeim94}).

The computation is based on the addition theorem of the second
order (see e.g. \cite{mu83})
\begin{eqnarray}
&&\theta[\varepsilon]({\bf x}|\tau)\theta[\delta]({\bf
y}|\tau)\label{add}\\ =&&\quad\sum_{\rho}
\theta\left[\begin{array}{c}{1\over2}(\varepsilon^{\prime}+
\delta^{\prime})+\rho
\\\varepsilon^{\prime\prime}+\delta^{\prime\prime}\end{array}\right]
({\bf x}+{\bf y}\big|2\tau)
\theta\left[\begin{array}{c}{1\over2}(\varepsilon^{\prime}-
\delta^{\prime})+\rho
\\\varepsilon^{\prime\prime}-\delta^{\prime\prime}\end{array}\right]
({\bf x}-{\bf y}\big|2\tau) ,\nonumber
\end{eqnarray}
where the summation
runs over $\rho =(0,0),(0,1),(1,0),(1,1)$ and
\begin{eqnarray}
&&\theta\left[\begin{array}{cc}\varepsilon_1^{\prime}&
\varepsilon_2^{\prime}
\\\varepsilon_1^{\prime\prime}&\varepsilon_2^{\prime\prime}
\end{array}\right]
\left({\bf
z}\big|\left(\begin{array}{cc}\tau&1\\1&\widetilde\tau\end{array}
\right)\right)\nonumber\\&&\quad={\rm e}^{-{1\over2}\pi i
\varepsilon_1^{\prime}\varepsilon_2^{\prime}}
\theta\left[\begin{array}{cc}\varepsilon_1^{\prime}&
\varepsilon_2^{\prime}
\\\varepsilon_1^{\prime\prime}+\varepsilon_2^{\prime}&
\varepsilon_2^{\prime\prime}+\varepsilon_1^{\prime}
\end{array}\right]
\left({\bf
z}\big|\left(\begin{array}{cc}\tau&0\\0&\widetilde\tau
\end{array}\right)\right)\label{decomp} \end{eqnarray}

We now show that under conditions $1^0$ and $2^0$ the argument of the
theta function  in (\ref{r1}-\ref{r5}) becomes $(x/2\omega,0)$.  The
condition $u_2=0$ or $\theta_1\left[{}_0^1{}_1^1\right]=0$ becomes the
condition (see Appendix)
\begin{equation}
{\sqrt{2}\vartheta^2_2\widetilde\vartheta_3\over\vartheta_3}+
\sqrt{\vartheta_3^2\widetilde\vartheta_3^2+
\vartheta_2^2\widetilde\vartheta_4^2 -
\vartheta_4^2\widetilde\vartheta_2^2}=0,\label{red}
\end{equation}
where $\vartheta_i=\vartheta_i(0;4\tau_{11})$, $
\widetilde\vartheta_i=\vartheta_i(0;4\tau_{22}),\;i=2,3,4.$ The
condition
(\ref{red}) is equivalent to the relations between Jacobi theta
constants
\begin{equation}
\tilde\vartheta_2^2=\tilde\vartheta_3^2{\vartheta_4^2\over
\vartheta_3^2}
\left(1-4{\vartheta_2^4\over\vartheta_3^4}\right),\;
\tilde\vartheta_4^2=\tilde\vartheta_3^2{\vartheta_2^2\over
\vartheta_3^2}
\left(1-4{\vartheta_4^4\over\vartheta_3^4}\right).
\label{reduc}
\end{equation}
Using the explicit expressions for the branching points (see Table 1)
we find after substitution of the corresponding theta constants (see
Appendix) in (\ref{wind})
\[
u_1={\sqrt{z_{\beta}-z_{\alpha}}\vartheta_3\over 2\pi
\vartheta_2\sqrt{4\vartheta_2^4-3\vartheta_3^4}}.
\]
Chosing
$z_{\beta,\alpha}=e_3+2e_1\pm2\sqrt{(e_1-e_3)(7e_1+2e_3)}$  we find
after the transformation $E=BAB$ (see e.g.\cite{ba55}),
$u_1=1/2\omega=\sqrt{e_1-e_3}/\pi\vartheta_3^2$, in  accordance with
\cite{ba55}.

The same arguments permit us to compute (\ref{qq1},\ref{qq2}) in terms
of elliptic functions. Let us consider the theta function entering into
the expression for (\ref{qq1}).  Applying the addition theorem
(\ref{add}) and putting $w_2=0$ according to condition $2^0$ and using
the expressions for theta constants we find
\begin{eqnarray}
&&{\fCA\left({x-x_0\over
2\omega},0\big|\tau\right)\over\fBA\left({x-x_0\over
2\omega},0\big|\tau\right)}=C_1{\vartheta_2\left({x-x_0\over
2\omega}\right)\over \vartheta_3\left({x-x_0\over
2\omega}\right)\vartheta_1^2\left({x-x_0\over
2\omega}\right)}\label{t0101}\\
&&\times\left[3\vartheta_4^2\vartheta_3^2\left({x-x_0\over
2\omega}\right) +\vartheta_2^2\vartheta_1^2\left({x-x_0\over
2\omega}\right)+\sqrt{\vartheta_3^4-4\vartheta_4^4}
\vartheta_1^2\left({x-x_0\over
2\omega}\right)\right]\nonumber\\
&&{\fAC\left({x-x_0\over
2\omega},0\big|\tau\right)\over\fBA\left({x-x_0\over
2\omega},0\big|\tau\right)}=C_2{\vartheta_4\left({x-x_0\over
2\omega}\right)\over \vartheta_3\left({x-x_0\over
2\omega}\right)\vartheta_1^2\left({x-x_0\over
2\omega}\right)}\label{t1010}\\
&&\times\left[3\vartheta_2^2\vartheta_3^2\left({x-x_0\over
2\omega}\right) -\vartheta_4^2\vartheta_1^2\left({x-x_0\over
2\omega}\right)+\sqrt{\vartheta_3^4-4\vartheta_2^4}
\vartheta_1^2\left({x-x_0\over
2\omega}\right)\right],\nonumber\end{eqnarray}
where $C_{1,2}$ can be computed using the theta constants from the
Appendix,
\begin{eqnarray}
C_2&=&{i\sqrt{2}\over2}{\vartheta_2^{5/2}\over\sqrt[4]{s}
\sqrt{\vartheta_2^4-\vartheta_4^4+\vartheta_2^2s}
(\vartheta_2^2+s)},\label{c1}\\
C_1&=&{\sqrt{2}\over2}(-1)^{1/4}{\vartheta_2^4\over
\vartheta_4^{3/2}\sqrt[4]{t}\sqrt{\vartheta_2^4-\vartheta_4^4-
\vartheta_4^2t}(t-\vartheta_4^2)},\nonumber
\end{eqnarray}
where $s=\sqrt{\vartheta_3^4-4\vartheta_4^4},t=
\sqrt{\vartheta_3^4-4\vartheta_2^4}$
One can see that substituting (\ref{t0101}, \ref{t1010})
and (\ref{q1},\ref{q2}) performing the transformation $BAB$ (see,
e.g.  \cite{ba55}) and setting $x_0=\omega_3$ we obtain the elliptic
solutions (\ref{ee1},\ref{ee2}). In the same way the another elliptic
solution associated with two-gap Treibich-Verdier potential
(\ref{ee3},\ref{ee4}) can be obtained from the theta functional
solutions.

\section{Acknowledgements} One of the authors (NAK) acknowledges
support from EEC grant ERB-CIPA-CT-92-0473 the other (VZE) acknowledges
support from the Royal Society under an ex-quota fellowship grant.
\newpage\appendix

\section {Theta Constants}
We denote  the  Jacobi theta constants by
$\vartheta_j=\vartheta_j\left(0|2^p\tau_{11}\right)$,
$\widetilde{\vartheta}_j=\vartheta_j\left(0|2^p\tau_{22}\right)$,
$j=2,3,4$.

\noindent
{\bf p=1:} Let $\tau=\left(\begin{array}{ll}\tau_{11}&{1\over2}\\
{1\over2}&\tau_{22}\end{array}\right)$.
Then
\begin{eqnarray}
&&\theta\ma{1}{0}{0}{0}=\theta\ma{1}{0}{0}{1}=
(2\vartheta_2\vartheta_3\widetilde{\vartheta}_3
\widetilde{\vartheta}_4
)^{1/2},\, \theta\ma{0}{1}{1}{0}=\theta\ma{0}{1}{0}{0}=
(2\vartheta_3\vartheta_4\widetilde{\vartheta}_2
\widetilde{\vartheta}_3
)^{1/2},\nonumber\\&&\hskip 28mm
\theta\ma{1}{1}{0}{0}=-i\theta\ma{1}{1}{1}{1}=
(2\vartheta_2\vartheta_4\widetilde{\vartheta}_2
\widetilde{\vartheta}_4
)^{1/2},\nonumber\\ &&\theta\ma{0}{0}{0}{0}=
(\vartheta_3^2\widetilde{\vartheta}_3^2+\vartheta_2^2
\widetilde{\vartheta}_4^2+\vartheta_4^2\widetilde{\vartheta}_2^2)
^{1/2},\, \theta\ma{0}{0}{1}{1}=
(\vartheta_3^2\widetilde{\vartheta}_3^2-\vartheta_2^2
\widetilde{\vartheta}_4^2-\vartheta_4^2\widetilde{\vartheta}_2^2)
^{1/2},\nonumber\\ &&\theta\ma{0}{0}{1}{0}=
(\vartheta_3^2\widetilde{\vartheta}_3^2-\vartheta_2^2
\widetilde{\vartheta}_4^2+\vartheta_4^2\widetilde{\vartheta}_2^2)
^{1/2},\, \theta\ma{0}{0}{0}{1}=
(\vartheta_3^2\widetilde{\vartheta}_3^2+\vartheta_2^2
\widetilde{\vartheta}_4^2-\vartheta_4^2\widetilde{\vartheta}_2^2)
^{1/2},\nonumber
\end{eqnarray}
\begin{eqnarray}
&&\theta_1\ma{1}{1}{1}{0}=-{\pi}\theta\ma{1}{1}{0}{0}
\vartheta_3^2,\quad \theta_2\ma{1}{1}{1}{0}=
-i\pi\theta\ma{1}{1}{1}{1}
\widetilde{\vartheta}_3^2,\nonumber\\
&&\theta_1\ma{1}{1}{0}{1}=-i\pi\theta\ma{1}{1}{0}{0}
\vartheta_3^2,\quad \theta_2\ma{1}{1}{0}{1}=
-\pi\theta\ma{1}{1}{0}{0}
\widetilde{\vartheta}_3^2,\nonumber\\
&&\theta_1\ma{0}{1}{0}{1}=-i\pi\theta\ma{0}{1}{0}{0}
\vartheta_2^2,\quad \theta_2\ma{0}{1}{0}{1}=
-\pi\theta\ma{0}{1}{0}{1}
\widetilde{\vartheta}_4^2,\nonumber\\
&&\theta_1\ma{0}{1}{1}{1}=i\pi\theta\ma{0}{1}{1}{0}
\vartheta_2^2,\quad
\theta_2\ma{0}{1}{1}{1}=-\pi\theta\ma{0}{1}{1}{0}
\widetilde{\vartheta}_4^2,\nonumber\\
&&\theta_1\ma{1}{0}{1}{1}=-\pi\theta\ma{1}{0}{0}{1}
\vartheta_4^2,\quad
\theta_2\ma{1}{0}{1}{1}=i\pi\theta\ma{1}{0}{0}{1}
\widetilde{\vartheta}_2^2,\nonumber\\
&&\theta_1\ma{1}{0}{1}{0}=-\pi\theta\ma{1}{0}{0}{0}
\vartheta_4^2,\quad
\theta_2\ma{1}{0}{1}{0}=-i\pi\theta\ma{1}{0}{0}{0}
\widetilde{\vartheta}_2^2.\nonumber\end{eqnarray}

\noindent
{\bf p=2:} Let $\tau=\left(\begin{array}{ll}
\tau_{11}&{1\over4}\\
{1\over4}&\tau_{22}\end{array}\right)$ and denote
$X=\vartheta_3\widetilde{\vartheta}_3$,
$Y=\vartheta_2\widetilde{\vartheta}_4$,
$Z=\vartheta_4\widetilde{\vartheta}_2$,
$A=-X^2+Y^2+Z^2$, $B=X^2-Y^2+Z^2$, $C=X^2+Y^2-Z^2$,
$D=A+B+C$.
Then the following formulae hold:

\begin{eqnarray}
&&\theta\ma{0}{0}{0}{0}=X+Y+Z,\quad\theta\ma{0}{0}{0}{1}=X+Y-Z,
\nonumber\\
&&\theta\ma{0}{0}{1}{0}=X-Y+Z,\quad\theta\ma{0}{0}{1}{1}=X-Y-Z,
\nonumber\\
&&\theta^2\ma{1}{0}{0}{0}=2\sqrt{2}(XY)^{1/2}(D^{1/2}+\sqrt{2}Z),\,
\theta^2\ma{1}{0}{0}{1}=2\sqrt{2}(XY)^{1/2}(D^{1/2}-\sqrt{2}Z),
\nonumber\\
&&\theta^2\ma{0}{1}{0}{0}=2\sqrt{2}(XZ)^{1/2}(D^{1/2}+\sqrt{2}Y),\,
\theta^2\ma{0}{1}{1}{0}=2\sqrt{2}(XZ)^{1/2}(D^{1/2}-\sqrt{2}Y),
\nonumber\\
&&\theta^2\ma{1}{1}{0}{0}=2\sqrt{2}(YZ)^{1/2}(D^{1/2}+\sqrt{2}X),
\theta^2\ma{1}{1}{1}{1}=2\sqrt{2}(YZ)^{1/2}(D^{1/2}-\sqrt{2}X).
\nonumber\end{eqnarray}

\begin{eqnarray} &&\theta_1\ma{1}{0}{1}{0}={ -\pi}(2XY)^{1/4}
(\vartheta_4^2B^{1/2}+\sqrt{2}\vartheta_3^2Z)
(D^{1/2}+\sqrt{2}Z)^{-1/2},\nonumber\\
&&\theta_2\ma{1}{0}{1}{0}=-{i\pi}(2XY)^{1/4}
(\widetilde{\vartheta}_2^2B^{1/2}+\sqrt{2}
\widetilde{\vartheta}_3^2Z)
(D^{1/2}+\sqrt{2}Z)^{-1/2},\nonumber\\
&&\theta_1\ma{0}{1}{0}{1}={{-i}\pi}(2XZ)^{1/4}
({\vartheta}_2^2C^{1/2}+\sqrt{2}{\vartheta}_3^2Y)
(D^{1/2}+\sqrt{2}Y)^{-1/2},\nonumber\\
&&\theta_2\ma{0}{1}{0}{1}=-{\pi}(2XZ)^{1/4}
(\widetilde{\vartheta}_4^2C^{1/2}+\sqrt{2}
\widetilde{\vartheta}_3^2Y)
(D^{1/2}+\sqrt{2}Y)^{-1/2},\nonumber\\ &&
\theta_1\ma{1}{1}{0}{1}={
-i\pi}(2ZY)^{1/4} ({\vartheta}_3^2C^{1/2}+
\sqrt{2}{\vartheta}_2^2X)
 (D^{1/2}+\sqrt{2}X)^{-1/2},\nonumber\\
&&\theta_2\ma{1}{1}{0}{1}=-{\pi}(2ZY)^{1/4}
(\widetilde{\vartheta}_3^2C^{1/2}+\sqrt{2}
\widetilde{\vartheta}_4^2X)
(D^{1/2}+\sqrt{2}X)^{-1/2},\nonumber\\
&&\theta_1\ma{1}{1}{1}{0}=-{\pi}(2ZY)^{1/4}
({\vartheta}_3^2B^{1/2}+\sqrt{2}{\vartheta}_4^2X)
 (D^{1/2}+\sqrt{2}X)^{-1/2},\nonumber\\
&&\theta_2\ma{1}{1}{1}{0}={-\pi}(2ZY)^{1/4}
(\widetilde{\vartheta}_3^2B^{1/2}+\sqrt{2}
\widetilde{\vartheta}_4^2X)
(D^{1/2}+\sqrt{2}X)^{-1/2},\nonumber\\
&&\theta_1\ma{1}{0}{1}{1}={{-i\pi}}(2XY)^{1/4}
(\vartheta_4^2A^{1/2}-\sqrt{2}i\vartheta_2^2Z)
 (D^{1/2}+\sqrt{2}Z)^{-1/2},\nonumber\\ &&
\theta_2\ma{1}{0}{1}{1}=-{
i\pi}(2XY)^{1/4}
(\widetilde{\vartheta}_2^2A^{1/2}-\sqrt{2}i
\widetilde{\vartheta}_3^2Z)
(D^{1/2}+\sqrt{2}Z)^{-1/2},\nonumber\\
&&\theta_1\ma{0}{1}{1}{1}={\pi}(2XZ)^{1/4}
({\vartheta}_2^2A^{1/2}-\sqrt{2}i{\vartheta}_3^2Y)
(D^{1/2}+\sqrt{2}Y)^{-1/2},\nonumber\\
&&\theta_2\ma{0}{1}{1}{1}=-i\pi(2XZ)^{1/4}
(\widetilde{\vartheta}_4^2A^{1/2}-\sqrt{2}i
\widetilde{\vartheta}_3^2Y)
(D^{1/2}+\sqrt{2}Y)^{-1/2}.\nonumber\end{eqnarray}

\end{document}